\documentclass[runningheads]{llncs}
\usepackage[T1]{fontenc}
\usepackage{graphicx,verbatim}
\usepackage{microtype}
\usepackage{subfigure}
\usepackage{graphicx}
\usepackage{caption}
\usepackage{subcaption}
\usepackage{amsmath}
\usepackage{amssymb}
\usepackage{bm}
\usepackage{booktabs}
\usepackage{multirow}
\usepackage{tabularx}
\usepackage{makecell}
\usepackage{marvosym}

\begin{document}
\title{UniSegDiff: Boosting Unified Lesion Segmentation via a Staged Diffusion Model}
\author{Yilong Hu\inst{1} \and
Shijie Chang\inst{1} \and
Lihe Zhang\inst{1} \textsuperscript{(\Letter)} \and
Feng Tian\inst{2} \and
Weibing Sun\inst{2} \and
Huchuan Lu\inst{1}
}

\institute{Dalian University of Technology, Dalian, China \\
\email{zhanglihe@dlut.edu.cn} \and
Department of Urology, Affiliated Zhongshan Hospital of Dalian University, Dalian, China
}

\maketitle              
\begin{abstract}
The Diffusion Probabilistic Model (DPM) has demonstrated remarkable performance across a variety of generative tasks. The inherent randomness in diffusion models helps address issues such as blurring at the edges of medical images and labels, positioning Diffusion Probabilistic Models (DPMs) as a promising approach for lesion segmentation. However, we find that the current training and inference strategies of diffusion models result in an uneven distribution of attention across different timesteps, leading to longer training times and suboptimal solutions. To this end, we propose UniSegDiff, a novel diffusion model framework designed to address lesion segmentation in a unified manner across multiple modalities and organs. This framework introduces a staged training and inference approach, dynamically adjusting the prediction targets at different stages, forcing the model to maintain high attention across all timesteps, and achieves unified lesion segmentation through pre-training the feature extraction network for segmentation. We evaluate performance on six different organs across various imaging modalities. Comprehensive experimental results demonstrate that UniSegDiff significantly outperforms previous state-of-the-art (SOTA) approaches. The code is available at https://github.com/HUYILONG-Z/UniSegDiff.


\end{abstract}
\section{Introduction}

Lesion segmentation is a critical task in medical image analysis. However, existing neural network models are typically designed for specific imaging modalities and lesion tasks~\cite{single,single2,single3,egenet}, which limits their broader applicability. Therefore, developing a unified model capable of handling multiple imaging modalities and lesion types is essential. In medical imaging, boundary ambiguity often arises in both images and labels~\cite{ambiguous}. To address this, we use Diffusion Probabilistic Models (DPMs)~\cite{ddpm} for medical lesion segmentation, as they incorporate randomness in modeling and can capture complex distributions. However, we observed that directly applying diffusion models to lesion segmentation tasks leads to longer convergence times, inference times, and suboptimal results, due to the uneven attention distribution across different timesteps. Through an in-depth analysis of the characteristics exhibited by diffusion models during training, we identified the root cause of the issue and developed a targeted staged diffusion framework, which was then applied to unified lesion segmentation.

\begin{figure}[h]
	\includegraphics[width=\textwidth]{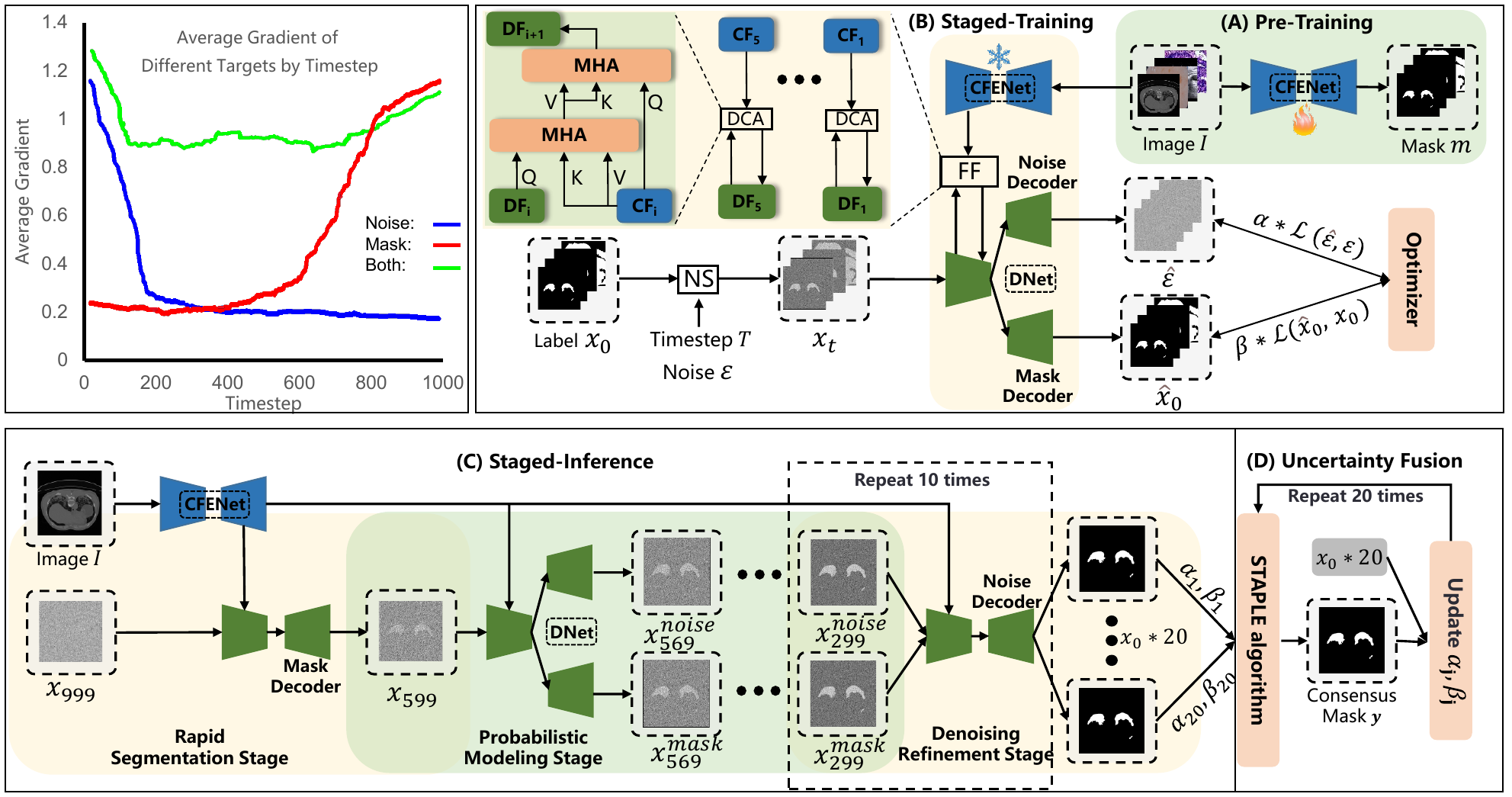}
	\caption{The top-left corner shows the relationship between the model's average gradient and timestep for different targets. (A) Pre-Training and (B) Stage-Training represent the training process of UniSegDiff, while (C) Staged-Inference and (D) Uncertainty Fusion illustrate the inference process. NS denotes Noise Schedule, FF stands for Features Fusion, DCA refers to Dual Cross Attention, and CF, DF represent Conditional Feature and Denoising Feature, respectively.}
	\label{fig:fig1}
\end{figure}

When diffusion models are applied to segmentation tasks, they typically consist of two parts: the conditional feature extraction network and the denoising network~\cite{segdiff}. The former encodes the image into conditional features to guide the latter in denoising training. The training and inference process is described as a Markov chain consisting of $T$ timesteps. As $t$ increases, the original mask $x_0$ is gradually corrupted by noise $\epsilon$ until it becomes pure Gaussian noise. The denoising network learns the ability to generate reconstructions by predicting $\epsilon$ or $x_0$ from the noisy mask $x_t$. Our observations indicate that predicting $\epsilon$ requires more training time to converge compared to predicting $x_0$. This is because, when the prediction target is $\epsilon$, the model finds it easier to learn the distribution of noise from noisy masks $x_t$ at larger timesteps than from those at smaller timesteps. As a result, the model tends to focus more on the latter (low-noise $x_t$). However, during inference, the model starts with pure Gaussian noise at the highest timestep and gradually denoises. The steps with larger timesteps are crucial in shaping the basic structure of the segmentation mask, which requires additional training for the model to converge at higher-noise timesteps. When the prediction target is $x_0$, the model tends to focus more on noisy masks at larger timesteps. Although the model can converge more quickly, it fails to model the noisy masks at smaller timesteps adequately, leading to poor performance. The upper-left part of Figure~\ref{fig:fig1} shows the average gradient distribution of the model across timesteps for different prediction targets, where higher values indicate greater attention from the model during that phase.


Moreover, applying diffusion models to unified lesion segmentation introduces new challenges. Different lesion images vary greatly in imaging modalities, lesion morphology, and other aspects, while the masks are simple binary images. This causes inevitable confusion of features from different lesions when using the lesion images as conditional guidance for denoising, leading to a mismatch between the conditional features and the denoising features.

To address these challenges, we propose a new framework called UniSegDiff. First, we divide different timesteps into three stages and dynamically set prediction targets: the Rapid Segmentation Stage (predicting $x_0$), the Probabilistic Modeling Stage (predicting both $x_0$ and $\epsilon$), and the Denoising Refinement Stage (predicting $\epsilon$), ensuring the model maintains high attention across all timesteps. Next, the conditional feature extraction network is pre-trained for segmentation on the unified lesion dataset and frozen during denoising training. This transforms lesion images from different modalities into distributions similar to the masks, reducing feature confusion between different lesions and better utilizing the conditional features to guide denoising. Finally, to fully leverage the inherent randomness modeled by the diffusion model, we use staged inference to quickly generate multiple segmentation results for fusion, obtaining the optimal solution. Our UniSegDiff achieves state-of-the-art performance on six lesion segmentation tasks across different modalities, as well as on the unified lesion segmentation task composed of these datasets.

    

\section{Approach}

\subsection{Overall Architecture}

The focus of this paper is on designing a diffusion framework for unified lesion segmentation, so our model architecture is kept simple, as shown in Figure~\ref{fig:fig1}. It consists of 2.5 UNet networks and a features fusion module. One UNet serves as the conditional feature extraction network (CFENet), while the remaining UNets function as the denoising network (DNet). During a single training step, CFENet extracts conditional features $CF_{i}\quad(i=1\sim5)$ from the input images. The encoder of DNet takes the noisy masks $x_t$, added by the noise scheduler, as input and progressively receives $CF_{i}$ as conditional guidance. Finally, the two decoders of DNet separately learn to model the $\epsilon$ and the $x_0$.

\subsection{Train and Infer Stage}

UniSegDiff divide the training process into three stages, each with distinct primary prediction objectives designed to ensure the neural network maintains high attention throughout all training steps. As shown in part (B) of Figure~\ref{fig:fig1}, in the Rapid Segmentation Stage ($599 < t$), the primary prediction target is the original mask $x_0$, as predicting the noise $\epsilon$ distribution is much simpler than predicting the original mask $x_0$ at this stage. Additionally, since the distribution difference of noisy masks $x_t$ at different time steps is minimal during this phase, all time steps are set to the maximum value ($t = 999$). Surprisingly, this not only accelerated the convergence speed but also improved segmentation accuracy. In the Probabilistic Modeling Stage ($299 < t \le 599$), the noise and mask information are more balanced, allowing the diffusion model to fully utilize its learning capability. At this stage, both prediction targets are given equal weight. In the Denoising Refinement Stage ($t \le 299$), the primary prediction target is the noise $\epsilon$. Similar to the Rapid Segmentation Stage, the distribution difference of noisy masks $x_t$ at different time steps is minimal, so all time steps are set to the minimum value ($t = 0$). The loss function of UniSegDiff is as follows:
\begin{equation}
    \mathcal{L}_{total} = \alpha \mathcal{L}_{n} + \beta (\mathcal{L}_{dice} + \mathcal{L}_{ce}).
\label{eq:eq1}
\end{equation}
The loss function consists of the noise prediction loss ($\mathcal{L}{n}$) and the original mask prediction loss ($\mathcal{L}{dice}$ + $\mathcal{L}_{ce}$), weighted accordingly. The weight coefficients $\alpha$ and $\beta$ are dynamically adjusted across different stages: in the Rapid Segmentation Stage ($\alpha : \beta$ = $1 : 3$), the Probabilistic Modeling Stage ($\alpha : \beta$ = $1 : 1$), and the Denoising Refinement Stage ($\alpha : \beta$ = $3 : 1$). This dynamic weighting scheme is consistent with our staged training approach.

The inference process is shown in part (C) of Figure~\ref{fig:fig1}, the initial time step of DNet is set to $t = 999$, with the input $X_{999}$ being pure Gaussian noise. After obtaining the conditional features $CF_{i}$ from CFENet, the mask prediction branch directly samples $X_{999}$ to $X_{599}$ in a single step. The subsequent sampling follows the DDIM method~\cite{ddim}, with a step interval of 30. After sampling $X_{599}$ ten times in each of the two decoder branches, the results are $X_{299}^{mask}$ and $X_{299}^{noise}$. Finally, the two noisy masks at $t = 299$ are each sampled ten times by the noise prediction branch, with each step directly sampling from $X_{299}$ to $X_{0}$. The 20 generated masks form a set of results, which are then prepared for subsequent uncertainty fusion. The entire sampling process is completed.

\subsection{Pre Train and Condition Injection}

To achieve unified lesion segmentation based on diffusion models, it is essential to eliminate the mismatch between conditional features and denoising features across different lesions. This requires the model to handle images from multiple modalities simultaneously and smoothly inject the conditional features of each lesion into the corresponding denoising features of the DNet. To this end, as shown in part (A) of Figure~\ref{fig:fig1}, we pre-train the CFENet on the unified dataset for the segmentation task and freeze it during the DNet training. This ensures that images from different modalities are transformed into distributions similar to the masks before being injected into DNet, narrowing the distribution gap between modalities and guiding DNet with the same set of features. This provides an appropriate prediction range for DNet, enabling it to refine and generate optimal results. During the stepwise injection of conditional features, we integrate them using the DCA (Dual Cross-Attention) module. The DCA module consists of two cascaded cross-attention blocks, with conditional features and noise mask features alternating as queries.

\subsection{Uncertainty Fusion}

For a lesion image $I$, we obtain a set of masks $z_j\quad(j = 1 \sim 20)$ through multiple samplings. To improve the model's accuracy and robustness, as shown in part (D) of Figure~\ref{fig:fig1}, we use the STAPLE~\cite{staple} algorithm to iteratively generate a consensus mask $y$. The confidence values $\alpha_j$ and $\beta_j$ for each $z_j$ are initialized to 0.9 and 0.1, respectively, representing the probabilities of correctly labeling the target and incorrectly labeling the background as the target. For each pixel $i$, the initial value of $y_i$ belongs to the target is set to $50\%$. The posterior probability of $y_i$ is calculated using the 20 masks $z_j$ through Equation.\ref{eq:eq2}. Then, using the maximum likelihood function from Equation.\ref{eq:eq3}, $\alpha_j$ and $\beta_j$ are updated based on new $y_i$. This update process is repeated 20 times to obtain the final consensus mask $y$.

\begin{equation}
    P(y_i = 1 | \{z_{ij}\}) = \frac{ \prod_{j=1}^{20} P(z_{ij} | y_i = 1, \alpha_j, \beta_j) P(y_i = 1) } { \sum_{y_i' \in \{0, 1\}} \prod_{j=1}^{20} P(z_{ij} | y_i', \alpha_j, \beta_j) P(y_i') }
    \label{eq:eq2}
\end{equation}

\begin{equation}
    \hat{\alpha}_j, \hat{\beta}_j = \frac{\sum_{i=1}^n P(y_i = \theta | \{z_{ij}\})}{\sum_{i=1}^n P(z_{ij} = 1 | y_i = \theta, \alpha_j, \beta_j)}, \quad \theta \in \{0, 1\}
    \label{eq:eq3}
\end{equation}

\section{Experiments}
\subsection{Datasets and Implementation Details}

We selected six publicly available and widely used lesion segmentation datasets from different organs and modalities to form a unified lesion segmentation dataset. The details are provided in Table~\ref{tab:ab1}. For colon polyp segmentation, we follow the  setting in Spider~\cite{spider}, combining five datasets to increase the challenge. Each dataset was randomly split into four equal parts for 4-fold cross-validation. For evaluation, we used two common metrics: mean Intersection over Union (mIoU) and mean Dice Similarity Coefficient (mDice). Detailed experimental setup, including the platform and hyperparameter settings, can be found in Table~\ref{tab:ab2}.

\begin{table}[h]
\centering
\fontsize{8}{10}\selectfont 
\begin{minipage}{0.59\linewidth} 
\centering
\caption{The dataset information of the six lesion segmentation tasks.}
\label{tab:ab1}
\begin{tabularx}{\linewidth}{cccc}
\toprule
\textbf{Task} & \textbf{Dataset} & \textbf{Modality} & \textbf{Images} \\
\midrule
Wet-AMD & AMD-SD~\cite{amdsd} & OCT & 3049 \\
Brain-Tumor & BTD~\cite{btd_1,btd_2} & MR-TI  & 3064 \\
Adenocarcinoma & EBHI-Seg~\cite{ebhi} & \makecell{Pathology\\image} & 795 \\
Colon Polyp & \makecell{Five datasets\\~\cite{ColonDB_polyp,endoscene_polyp,etis_polyp,kvasir_polyp,pranet_polyp}} & \makecell{Endoscopy\\image} & 2248 \\
Lung Infection & COVID-19~\cite{jun2020covid,medicalsegmentation_covid19} & CT & 1277 \\
Breast Lesion & BUSI~\cite{busi} & Ultrasound & 647 \\
\bottomrule
\end{tabularx}
\end{minipage}
\hfill
\begin{minipage}{0.39\linewidth} 
\centering
\caption{Implementation Details}
\label{tab:ab2}
\begin{tabularx}{\linewidth}{cc}
\toprule
\textbf{Category} & \textbf{Details} \\
\midrule
Framework & PyTorch \\
Hardware & 4 $\times$ 3090 GPUs \\
\makecell{Image\\Resolution} & $256 \times 256$ \\
Optimizer & AdamW \\
Lr Scheduler & \text{CosineAnnealingLR} \\
Initial Lr & $1e^{-4}$ \\
Total Epochs & 300 \\
Batch Size & 64 \\
\bottomrule
\end{tabularx}
\end{minipage}
\end{table}


\subsection{Evaluation}

\noindent\textbf{Comparison with State-of-the-Arts} To validate the effectiveness of UniSegDiff, we compared it with SOTA discriminative segmentation methods~\cite{unet,transunet,rollingunet,mednext,emcad} and diffusion-based segmentation methods~\cite{medsegdiffv2,sdseg,cdal} on both the unified lesion segmentation task and six individual lesion segmentation tasks. The quantitative results are presented in Table~\ref{tab:ab3}. UniSegDiff consistently outperforms all models across both single-task and unified tasks. In the unified lesion segmentation task, all models showed a significant performance decline. However, thanks to the pre-training of CFENet, UniSegDiff reduced the distribution gap between datasets, resulting in no noticeable performance drop in the unified lesion segmentation task. As a result, it outperformed other methods by a considerable margin.

\begin{table}[h!]
\centering
\fontsize{8}{3}\selectfont 
\caption{The quantitative comparisons across various lesion segmentation tasks. From left to right in Table~\ref{tab:ab3}, the six tasks are those listed in Table~\ref{tab:ab1}. The values following $\pm$ represent the standard deviation.}
\label{tab:ab3}
\begin{tabularx}{\linewidth}{>{\centering\arraybackslash}p{2cm} *{12}{>{\centering\arraybackslash}X}}
\toprule
\multirow{2}{*}{Methods} & \multicolumn{2}{c}{\textbf{WA}} & \multicolumn{2}{c}{\textbf{BT}} & \multicolumn{2}{c}{\textbf{ADC}} & \multicolumn{2}{c}{\textbf{CP}} & \multicolumn{2}{c}{\textbf{LI}} & \multicolumn{2}{c}{\textbf{BL}} \\
& mDice & mIoU & mDice & mIoU & mDice & mIoU & mDice & mIoU & mDice & mIoU & mDice & mIoU \\
\midrule
\multicolumn{13}{c}{\textbf{Individual Lesion Segmentation Tasks}} \\
\midrule
\multirow{2}{*}{UNet} 
& 86.8 & 77.6 & 81.8 & 73.2 & 91.6 & 85.2 & 82.6 & 75.5 & 73.1 & 65.4 & 74.9 & 65.5 \\
& $\pm0.51$ & $\pm0.69$ & $\pm0.8$ & $\pm0.63$ & $\pm0.47$ & $\pm0.56$ & $\pm0.44$ & $\pm0.52$ & $\pm1.83$ & $\pm1.64$ & $\pm0.72$ & $\pm0.82$ \\[1ex]

\multirow{2}{*}{TransUNet}
& 86.0 & 76.4 & 81.1 & 72.2 & 91.5 & 85.0 & 84.6 & 77.9 & 74.2 & 66.4 & 78.6 & 69.6 \\
& $\pm0.53$ & $\pm0.70$ & $\pm1.37$ & $\pm1.25$ & $\pm0.39$ & $\pm0.62$ & $\pm0.61$ & $\pm0.62$ & $\pm1.73$ & $\pm1.72$ & $\pm0.54$ & $\pm0.36$ \\[1ex]

\multirow{2}{*}{RollingUNet}
& 84.1 & 74.1 & 78.7 & 69.0 & 91.2 & 84.5 & 85.0 & 78.5 & 76.4 & 68.8 & 78.5 & 69.5 \\
& $\pm1.41$ & $\pm1.80$ & $\pm0.78$ & $\pm0.83$ & $\pm1.20$ & $\pm2.07$ & $\pm0.55$ & $\pm0.70$ & $\pm1.69$ & $\pm1.56$ & $\pm0.54$ & $\pm0.40$ \\[1ex]

\multirow{2}{*}{MedNeXt}
& 86.8 & 77.6 & 83.2 & 74.7 & 92.1 & 86.2 & 88.9 & 83.2 & 75.4 & 68.0 & 80.0 & 71.5 \\
& $\pm0.54$ & $\pm0.65$ & $\pm0.78$ & $\pm0.65$ & $\pm0.37$ & $\pm0.41$ & $\pm0.51$ & $\pm0.51$ & $\pm1.63$ & $\pm1.41$ & $\pm0.42$ & $\pm0.43$ \\[1ex]

\multirow{2}{*}{EMCAD}
& 84.3 & 73.9 & 82.8 & 74.1 & 93.0 & 87.2 & 86.9 & 81.3 & 67.4 & 59.9 & 78.5 & 69.4 \\
& $\pm0.39$ & $\pm0.49$ & $\pm0.52$ & $\pm0.69$ & $\pm0.2$ & $\pm0.27$ & $\pm0.32$ & $\pm0.32$ & $\pm3.4$ & $\pm3.33$ & $\pm1.94$ & $\pm2.0$ \\[1ex]

\multirow{2}{*}{Medsegdiff-V2}
& 86.7 & 77.5 & 80.9 & 71.9 & 91.1 & 84.5 & 85.2 & 78.5 & 77.7 & 70.4 & 78.9 & 70.2 \\
& $\pm0.59$ & $\pm0.89$ & $\pm0.98$ & $\pm1.23$ & $\pm0.69$ & $\pm0.91$ & $\pm0.26$ & $\pm0.50$ & $\pm1.87$ & $\pm1.81$ & $\pm2.0$ & $\pm1.9$ \\[1ex]

\multirow{2}{*}{cDAL}
& 78.3 & 65.8 & 80.0 & 70.5 & 67.4 & 62.1 & 85.7 & 79.1 & 74.6 & 66.4 & 76.9 & 68.0 \\
& $\pm0.62$ & $\pm0.87$ & $\pm0.71$ & $\pm0.75$ & $\pm0.31$ & $\pm0.28$ & $\pm0.78$ & $\pm0.89$ & $\pm0.79$ & $\pm0.45$ & $\pm0.98$ & $\pm0.91$ \\[1ex]

\multirow{2}{*}{SDSeg}
& 85.1 & 75.2 & 81.2 & 72.4 & 91.3 & 84.9 & 86.6 & 80.1 & 76.6 & 69.2 & 78.7 & 70.1 \\
& $\pm0.5$ & $\pm0.57$ & $\pm0.8$ & $\pm0.7$ & $\pm0.41$ & $\pm0.45$ & $\pm0.66$ & $\pm0.75$ & $\pm1.11$ & $\pm1.07$ & $\pm1.86$ & $\pm1.66$ \\[1ex]

\multirow{2}{*}{UniSegDiff}
& \textbf{87.1} & \textbf{78.1} & \textbf{84.5} & \textbf{76.0} & \textbf{93.0} & \textbf{87.3} & \textbf{89.0} & \textbf{82.9} & \textbf{79.9} & \textbf{72.4} & \textbf{81.9} & \textbf{73.1} \\
& $\pm0.59$ & $\pm0.79$ & $\pm0.71$ & $\pm0.58$ & $\pm0.26$ & $\pm0.44$ & $\pm1.23$ & $\pm1.22$ & $\pm1.25$ & $\pm1.39$ & $\pm3.1$ & $\pm3.59$ \\[1ex]

\midrule
\multicolumn{13}{c}{\textbf{Unified Lesion Segmentation Task}} \\
\midrule
\multirow{2}{*}{UNet} 
& 84.5 & 74.2 & 80.0 & 71.2 & 90.2 & 83.1 & 79.6 & 71.1 & 64.5 & 56.0 & 71.1 & 61.3 \\
& $\pm0.73$ & $\pm0.87$ & $\pm0.85$ & $\pm0.87$ & $\pm0.57$ & $\pm0.78$ & $\pm0.23$ & $\pm0.25$ & $\pm1.46$ & $\pm1.44$ & $\pm1.2$ & $\pm1.23$ \\[1ex]

\multirow{2}{*}{TransUNet}
& 82.7 & 71.6 & 76.8 & 67.6 & 90.3 & 83.1 & 79.7 & 71.2 & 64.3 & 56.1 & 75.1 & 65.7 \\
& $\pm0.3$ & $\pm0.37$ & $\pm1.78$ & $\pm1.94$ & $\pm0.82$ & $\pm0.89$ & $\pm1.56$ & $\pm1.61$ & $\pm1.5$ & $\pm1.59$ & $\pm1.19$ & $\pm1.3$ \\[1ex]

\multirow{2}{*}{RollingUNet}
& 83.4 & 72.8 & 78.4 & 68.8 & 89.9 & 82.6 & 79.9 & 71.3 & 56.2 & 47.6 & 72.8 & 62.3 \\
& $\pm1.43$ & $\pm1.79$ & $\pm1.48$ & $\pm1.65$ & $\pm0.96$ & $\pm1.3$ & $\pm0.37$ & $\pm0.3$ & $\pm1.52$ & $\pm1.59$ & $\pm0.76$ & $\pm0.71$ \\[1ex]

\multirow{2}{*}{MedNeXt}
& 84.2 & 73.9 & 80.6 & 71.7 & 90.6 & 83.6 & 84.0 & 76.7 & 63.8 & 55.7 & 76.7 & 67.5 \\
& $\pm0.33$ & $\pm0.33$ & $\pm0.78$ & $\pm0.73$ & $\pm0.37$ & $\pm0.52$ & $\pm1.0$ & $\pm1.42$ & $\pm1.33$ & $\pm1.48$ & $\pm2.1$ & $\pm1.91$ \\[1ex]

\multirow{2}{*}{EMCAD}
& 83.5 & 72.7 & 82.9 & 74.2 & 91.6 & 85.9 & 86.9 & 81.3 & 65.2 & 57.7 & 77.5 & 68.6 \\
& $\pm0.64$ & $\pm0.87$ & $\pm0.66$ & $\pm0.72$ & $\pm0.27$ & $\pm0.47$ & $\pm0.68$ & $\pm0.89$ & $\pm3.1$ & $\pm2.9$ & $\pm0.84$ & $\pm0.86$ \\[1ex]

\multirow{2}{*}{Medsegdiff-V2}
& 83.5 & 72.9 & 78.0 & 68.7 & 91.2 & 84.6 & 78.1 & 69.9 & 57.7 & 49.2 & 73.8 & 64.7 \\
& $\pm0.71$ & $\pm0.97$ & $\pm1.55$ & $\pm1.69$ & $\pm0.6$ & $\pm0.76$ & $\pm1.4$ & $\pm1.37$ & $\pm1.53$ & $\pm1.52$ & $\pm0.66$ & $\pm0.67$ \\[1ex]

\multirow{2}{*}{cDAL}
& 76.7 & 64.0 & 73.8 & 64.4 & 67.3 & 62.1 & 82.8 & 75.6 & 65.9 & 57.6 & 75.5 & 66.6 \\
& $\pm0.63$ & $\pm0.69$ & $\pm0.83$ & $\pm0.99$ & $\pm1.26$ & $\pm1.23$ & $\pm0.94$ & $\pm0.97$ & $\pm1.38$ & $\pm1.36$ & $\pm1.84$ & $\pm1.72$ \\[1ex]

\multirow{2}{*}{SDSeg}
& 67.2 & 55.6 & 76.6 & 66.9 & 90.9 & 84.2 & 86.6 & 79.9 & 62.1 & 54.7 & 78.3 & 69.7 \\
& $\pm0.95$ & $\pm0.97$ & $\pm0.89$ & $\pm0.87$ & $\pm1.41$ & $\pm1.45$ & $\pm0.72$ & $\pm0.75$ & $\pm1.34$ & $\pm1.31$ & $\pm1.78$ & $\pm1.82$ \\[1ex]

\multirow{2}{*}{UniSegDiff}
& \textbf{86.8} & \textbf{77.6} & \textbf{83.3} & \textbf{74.7} & \textbf{92.0} & \textbf{85.9} & \textbf{88.4} & \textbf{82.4} & \textbf{79.4} & \textbf{71.9} & \textbf{79.5} & \textbf{70.4} \\
& $\pm0.5$ & $\pm0.72$ & $\pm0.29$ & $\pm0.18$ & $\pm0.28$ & $\pm0.3$ & $\pm0.71$ & $\pm0.96$ & $\pm1.25$ & $\pm1.3$ & $\pm1.32$ & $\pm1.29$ \\[1ex]

\bottomrule
\end{tabularx}
\end{table}

\noindent\textbf{Defect analysis} All methods showed a significant performance drop on the Lung Infection task during unified lesion segmentation. After examining the dataset, we found that this was due to a large number of masks being empty (approximately one-third of the dataset). We will clean the data and re-validate the results in future work.

\subsection{Ablation Study}

In this section, we examine how different denoising methods influence segmentation performance, as well as training and inference speed. We also analyze the impact of threshold selection in our proposed staged training method and the contribution of each component in the network. All experiments were conducted on the unified lesion segmentation task. Due to space limitations in the table, we only present the average values of the metrics across all datasets for the unified segmentation task, without showing the standard deviation.

\noindent\textbf{Effectiveness of denoising methods}
Table~\ref{tab:ab4} compares different denoising training strategies for the diffusion model. In traditional uniform denoising, predicting noise takes significantly more training epochs to converge. Direct original mask prediction accelerates convergence but still requires at least 100 inference steps for satisfactory results~\cite{ddim}. One-step denoising~\cite{onestep} achieves faster training and inference but performs better for original mask prediction than for noise, likely due to the network’s preference for distribution mapping. While fast, this method sacrifices accuracy. Our staged denoising strategy balances efficiency and accuracy: During training, dynamic prediction targets ensure high attention across all time steps, facilitating rapid convergence and fully leveraging the model’s capabilities. During inference, the rapid segmentation and denoising refinement stages perform single-step sampling, achieving accurate segmentation in as few as eleven steps (multiple refinements for mask fusion yield optimal results). This approach is at least 10 times faster than DDIM and 100 times faster than DDPM.

\begin{table}[h]
\centering
\fontsize{8}{10}\selectfont 
\begin{minipage}{0.55\linewidth} 
\centering
\caption{Ablation experiments of the denoising methods.}
\label{tab:ab4}
\begin{tabularx}{\linewidth}{>{\centering\arraybackslash}X >{\centering\arraybackslash}X >{\centering\arraybackslash}X >{\centering\arraybackslash}X | >{\centering\arraybackslash}X >{\centering\arraybackslash}X}
\toprule
\multirow{2}{*}{\makecell{Denoise\\Method}} & 
\multirow{2}{*}{Target} & 
\multirow{2}{*}{\makecell{Train\\Epoch}} & 
\multirow{2}{*}{\makecell{Infer\\Step}} & 
\multicolumn{2}{c}{\makecell{Unified Task}} \\
& & & & mDice & mIoU \\ 
\midrule
\multirow{2}{*}{Uniform} & noise & 1000 & 100 & 81.5 & 73.4 \\ 
& mask & 300 & 100 & 80.9 & 72.6 \\ 
\multirow{2}{*}{One-Step} & noise & 300 & 1 & 75.3 & 68.2 \\ 
& mask & 300 & 1 & 78.6 & 71.4 \\
Staged & both & 300 & 11 & \textbf{84.4} & \textbf{76.3} \\ 
\bottomrule
\end{tabularx}
\end{minipage}
\hfill
\begin{minipage}{0.43\linewidth} 
\centering
\caption{Ablation experiments of the threshold selection}
\label{tab:ab5}
\begin{tabularx}{\linewidth}{>{\centering\arraybackslash}X >{\centering\arraybackslash}X | >{\centering\arraybackslash}X >{\centering\arraybackslash}X}
\toprule
\multirow{2}{*}{\makecell{High\\threshold}} & 
\multirow{2}{*}{\makecell{Low\\threshold}} & 
\multicolumn{2}{c}{\makecell{Unified Task}} \\
& & mDice & mIoU \\
\midrule
700 &    & 82.1 & 74.0 \\ 
600 &    & 83.3 & 75.6 \\
500 &    & 82.7 & 74.7 \\
\hline
600 & 400    & 83.9 & 76.8 \\
600 & 300    & \textbf{84.4} & \textbf{76.3} \\ 
600 & 200    & 84.4 & 76.3 \\
\bottomrule
\end{tabularx}
\end{minipage}
\end{table}

\begin{table}[h]
\centering
\fontsize{8}{10}\selectfont 
\begin{minipage}{0.42\linewidth} 
\centering
\caption{An ablation experiments of each component.}
\label{tab:ab6}
\centering
\begin{tabularx}{\linewidth}{>{\centering\arraybackslash}X >{\centering\arraybackslash}X >{\centering\arraybackslash}X >{\centering\arraybackslash}X | >{\centering\arraybackslash}X}
\toprule
\multirow{3}{*}{Staged} & 
\multirow{3}{*}{Pre-Tra} & 
\multirow{3}{*}{DCA} & 
\multirow{3}{*}{Fusion} & 
Unified Task \\
& & & & mDice \\ 
\midrule
& & & & 77.0 \\ 
\checkmark  & & &   & 80.5 \\ 
\checkmark & \checkmark & &   & 83.8 \\
\checkmark & \checkmark & \checkmark &   & 84.4 \\ 
\checkmark & \checkmark & \checkmark & \checkmark   & \textbf{85.3} \\ 
\bottomrule
\end{tabularx}
\end{minipage}
\hfill
\begin{minipage}{0.57\linewidth} 
\centering
\caption{Comparison of training time, inference speed and inference Steps.}
\label{tab:ab7}
\begin{tabularx}{\linewidth}{>{\centering\arraybackslash}X >{\centering\arraybackslash}X >{\centering\arraybackslash}X >{\centering\arraybackslash}X | >{\centering\arraybackslash}X}
\toprule
\multirow{3}{*}{Methods} & 
\multirow{3}{*}{\makecell{Training\\Time\\(hours)}} & 
\multirow{3}{*}{\makecell{Inference\\Speed\\(samples/s)}} & 
\multirow{3}{*}{\makecell{inference\\Steps}} & 
Unified Task \\ 
& & & & mDice \\
\midrule
Medsegdiff & $\approx172$ & 0.24 & 100 & 77.1 \\
SDSeg & $\approx43$ & \textbf{13.3} & 1 &77.0 \\
cDAL & $\approx110$ & 1.18 & 60 & 73.7 \\
UniSegDiff & $\approx\textbf{25}$ & 8.95 & 11 & \textbf{84.4} \\
\bottomrule
\end{tabularx}
\end{minipage}
\end{table}

\noindent\textbf{Ablation Study on Threshold Selection}
The staged training and inference approach we propose is divided into three phases, with the threshold selection between phases being critical. Table~\ref{tab:ab5} presents detailed ablation experiments. In the experiments, the high threshold was first set to $t = 600$, and then, with the high threshold fixed, different low threshold values were tested. Ultimately, it was found that the optimal low threshold is $t = 300$.


\noindent\textbf{Effectiveness of each component}
Table~\ref{tab:ab6} presents the ablation experiments for each component proposed in UniSegDiff. The baseline uses uniform sampling to predict $x_0$ during training. Clearly, while the staged training approach improves segmentation performance, pre-training CFENet for segmentation significantly enhances the model’s accuracy in unified lesion segmentation. The DCA module further facilitates feature fusion between CFENet and DNet, while uncertainty fusion leverages the randomness of the diffusion model to further enhance the accuracy and robustness of the segmentation results.

\noindent\textbf{Comparison of time efficiency}
Table~\ref{tab:ab7} presents the efficiency evaluation results for MedSegDiff-V2, SDSeg, cDAL, and UniSegDiff in the unified lesion segmentation task. To ensure a fair comparison, all models were trained on the same server. The results show that UniSegDiff significantly reduced training time and is much faster during inference compared to MedSegDiff-V2 and cDAL. Although it is slower than SDSeg, which samples only once, UniSegDiff achieves segmentation accuracy far superior to other diffusion-based models.


\section{Conclusion and Future Work}

In this paper, we investigate the characteristics of applying diffusion models to segmentation tasks. Through analysis, we propose a staged diffusion framework for unified lesion segmentation tasks, which includes tailored training strategies, inference methods, and model architecture. To enhance alignment across different types of lesion data, we pre-train the conditional feature extraction network as a segmentation model, significantly improving both inference speed and segmentation accuracy. Our method achieves state-of-the-art performance across multiple lesion segmentation benchmarks. Future work will focus on expanding our dataset to cover more lesion types and extending our approach into a unified framework that supports both 2D and 3D lesion data, with the goal of achieving comprehensive segmentation for all lesion types.

\subsubsection{\ackname}
This work was supported by Dalian Science and Technology Innovation Foundation under Grant 2023JJ12GX015, and by the National Natural Science Foundation of China under Grant 62276046 and 62431004.

\subsubsection{\discintname}
The authors have no competing interests to declare that are
relevant to the content of this article

%
%
%
\bibliographystyle{splncs04}
\bibliography{mybibliography}

\begin{thebibliography}{10}
\providecommand{\url}[1]{\texttt{#1}}
\providecommand{\urlprefix}{URL }
\providecommand{\doi}[1]{https://doi.org/#1}

\bibitem{medicalsegmentation_covid19}
Covid-19 ct lung and infection segmentation dataset. \url{https://medicalsegmentation.com/covid19/} (2020), accessed: June 25, 2025

\bibitem{busi}
Al-Dhabyani, W., Gomaa, M., Khaled, H., Fahmy, A.: Dataset of breast ultrasound images. Data in brief  \textbf{28},  104863 (2020)

\bibitem{segdiff}
Amit, T., Shaharbany, T., Nachmani, E., Wolf, L.: Segdiff: Image segmentation with diffusion probabilistic models. arXiv preprint arXiv:2112.00390  (2021)

\bibitem{transunet}
Chen, J., Lu, Y., Yu, Q., Luo, X., Adeli, E., Wang, Y., Lu, L., Yuille, A.L., Zhou, Y.: Transunet: Transformers make strong encoders for medical image segmentation. arXiv preprint arXiv:2102.04306  (2021)

\bibitem{btd_1}
Cheng, J., Huang, W., Cao, S., Yang, R., Yang, W., Yun, Z., Wang, Z., Feng, Q.: Enhanced performance of brain tumor classification via tumor region augmentation and partition. PloS one  \textbf{10}(10),  e0140381 (2015)

\bibitem{btd_2}
Cheng, J., Yang, W., Huang, M., Huang, W., Jiang, J., Zhou, Y., Yang, R., Zhao, J., Feng, Y., Feng, Q., et~al.: Retrieval of brain tumors by adaptive spatial pooling and fisher vector representation. PloS one  \textbf{11}(6),  e0157112 (2016)

\bibitem{pranet_polyp}
Fan, D.P., Ji, G.P., Zhou, T., Chen, G., Fu, H., Shen, J., Shao, L.: Pranet: Parallel reverse attention network for polyp segmentation. In: MICCAI. pp. 263--273. Springer (2020)

\bibitem{cdal}
Hejrati, B., Banerjee, S., Glide-Hurst, C., Dong, M.: Conditional diffusion model with spatial attention and latent embedding for medical image segmentation. In: International Conference on Medical Image Computing and Computer-Assisted Intervention. pp. 202--212. Springer (2024)

\bibitem{ddpm}
Ho, J., Jain, A., Abbeel, P.: Denoising diffusion probabilistic models. Advances in neural information processing systems  \textbf{33},  6840--6851 (2020)

\bibitem{amdsd}
Hu, Y., Gao, Y., Gao, W., Luo, W., Yang, Z., Xiong, F., Chen, Z., Lin, Y., Xia, X., Yin, X., et~al.: Amd-sd: An optical coherence tomography image dataset for wet amd lesions segmentation. Scientific Data  \textbf{11}(1), ~1014 (2024)

\bibitem{kvasir_polyp}
Jha, D., Smedsrud, P.H., Riegler, M.A., Halvorsen, P., De~Lange, T., Johansen, D., Johansen, H.D.: Kvasir-seg: A segmented polyp dataset. In: MultiMedia modeling: 26th international conference, MMM 2020, Daejeon, South Korea, January 5--8, 2020, proceedings, part II 26. pp. 451--462. Springer (2020)

\bibitem{jun2020covid}
Jun, M., Cheng, G., Yixin, W., Xingle, A., Jiantao, G., Ziqi, Y., Minqing, Z., Xin, L., Xueyuan, D., Shucheng, C., et~al.: Covid-19 ct lung and infection segmentation dataset. (No Title)  (2020)

\bibitem{single}
Kim, R.Y., Oke, J.L., Pickup, L.C., Munden, R.F., Dotson, T.L., Bellinger, C.R., Cohen, A., Simoff, M.J., Massion, P.P., Filippini, C., et~al.: Artificial intelligence tool for assessment of indeterminate pulmonary nodules detected with ct. Radiology  \textbf{304}(3),  683--691 (2022)

\bibitem{sdseg}
Lin, T., Chen, Z., Yan, Z., Yu, W., Zheng, F.: Stable diffusion segmentation for biomedical images with single-step reverse process. In: International Conference on Medical Image Computing and Computer-Assisted Intervention. pp. 656--666. Springer (2024)

\bibitem{rollingunet}
Liu, Y., Zhu, H., Liu, M., Yu, H., Chen, Z., Gao, J.: Rolling-unet: Revitalizing mlp’s ability to efficiently extract long-distance dependencies for medical image segmentation. In: Proceedings of the AAAI Conference on Artificial Intelligence. vol.~38, pp. 3819--3827 (2024)

\bibitem{onestep}
Parmar, G., Park, T., Narasimhan, S., Zhu, J.Y.: One-step image translation with text-to-image models. arXiv preprint arXiv:2403.12036  (2024)

\bibitem{ambiguous}
Rahman, A., Valanarasu, J.M.J., Hacihaliloglu, I., Patel, V.M.: Ambiguous medical image segmentation using diffusion models. In: Proceedings of the IEEE/CVF conference on computer vision and pattern recognition. pp. 11536--11546 (2023)

\bibitem{emcad}
Rahman, M.M., Munir, M., Marculescu, R.: Emcad: Efficient multi-scale convolutional attention decoding for medical image segmentation. In: Proceedings of the IEEE/CVF Conference on Computer Vision and Pattern Recognition. pp. 11769--11779 (2024)

\bibitem{unet}
Ronneberger, O., Fischer, P., Brox, T.: U-net: Convolutional networks for biomedical image segmentation. In: Medical image computing and computer-assisted intervention--MICCAI 2015: 18th international conference, Munich, Germany, October 5-9, 2015, proceedings, part III 18. pp. 234--241. Springer (2015)

\bibitem{mednext}
Roy, S., Koehler, G., Ulrich, C., Baumgartner, M., Petersen, J., Isensee, F., Jaeger, P.F., Maier-Hein, K.H.: Mednext: transformer-driven scaling of convnets for medical image segmentation. In: International Conference on Medical Image Computing and Computer-Assisted Intervention. pp. 405--415. Springer (2023)

\bibitem{egenet}
Ruan, J., Xie, M., Gao, J., Liu, T., Fu, Y.: Ege-unet: an efficient group enhanced unet for skin lesion segmentation. In: International conference on medical image computing and computer-assisted intervention. pp. 481--490. Springer (2023)

\bibitem{ebhi}
Shi, L., Li, X., Hu, W., Chen, H., Chen, J., Fan, Z., Gao, M., Jing, Y., Lu, G., Ma, D., et~al.: Ebhi-seg: A novel enteroscope biopsy histopathological hematoxylin and eosin image dataset for image segmentation tasks. Frontiers in Medicine  \textbf{10},  1114673 (2023)

\bibitem{etis_polyp}
Silva, J., Histace, A., Romain, O., Dray, X., Granado, B.: Toward embedded detection of polyps in wce images for early diagnosis of colorectal cancer. International journal of computer assisted radiology and surgery  \textbf{9},  283--293 (2014)

\bibitem{ddim}
Song, J., Meng, C., Ermon, S.: Denoising diffusion implicit models. arXiv preprint arXiv:2010.02502  (2020)

\bibitem{ColonDB_polyp}
Tajbakhsh, N., Gurudu, S.R., Liang, J.: Automated polyp detection in colonoscopy videos using shape and context information. IEEE TMI  \textbf{35}(2),  630--644 (2015)

\bibitem{endoscene_polyp}
V{\'a}zquez, D., Bernal, J., S{\'a}nchez, F.J., Fern{\'a}ndez-Esparrach, G., L{\'o}pez, A.M., Romero, A., Drozdzal, M., Courville, A.: A benchmark for endoluminal scene segmentation of colonoscopy images. Journal of healthcare engineering  \textbf{2017}(1),  4037190 (2017)

\bibitem{staple}
Warfield, S.K., Zou, K.H., Wells, W.M.: Simultaneous truth and performance level estimation (staple): an algorithm for the validation of image segmentation. IEEE transactions on medical imaging  \textbf{23}(7),  903--921 (2004)

\bibitem{medsegdiffv2}
Wu, J., Ji, W., Fu, H., Xu, M., Jin, Y., Xu, Y.: Medsegdiff-v2: Diffusion-based medical image segmentation with transformer. In: Proceedings of the AAAI Conference on Artificial Intelligence. vol.~38, pp. 6030--6038 (2024)

\bibitem{spider}
Zhao, X., Pang, Y., Ji, W., Sheng, B., Zuo, J., Zhang, L., Lu, H.: Spider: A unified framework for context-dependent concept understanding. arXiv preprint arXiv:2405.01002  (2024)

\bibitem{single3}
Zhu, L., Xue, Z., Jin, Z., Liu, X., He, J., Liu, Z., Yu, L.: Make-a-volume: Leveraging latent diffusion models for cross-modality 3d brain mri synthesis. In: International Conference on Medical Image Computing and Computer-Assisted Intervention. pp. 592--601. Springer (2023)

\bibitem{single2}
Zhu, Z., Xia, Y., Xie, L., Fishman, E.K., Yuille, A.L.: Multi-scale coarse-to-fine segmentation for screening pancreatic ductal adenocarcinoma. In: Medical Image Computing and Computer Assisted Intervention--MICCAI 2019: 22nd International Conference, Shenzhen, China, October 13--17, 2019, Proceedings, Part VI 22. pp. 3--12. Springer (2019)

\end{thebibliography}
%




\end{document}